\documentclass[12pt]{article}

\usepackage{scicite}
\usepackage{times}

\newenvironment{sciabstract}{%
\begin{quote} \bf}
{\end{quote}}

\newcounter{lastnote}

\usepackage[utf8]{inputenc}
\usepackage{geometry}
\usepackage{authblk} 
\usepackage{braket} 
\usepackage{tabularx}
\usepackage{graphicx}
\usepackage{xcolor}
\usepackage{tikz}
\usetikzlibrary{calc}
\usetikzlibrary{matrix}
\usepackage{caption}
\usepackage{subcaption}
\usepackage{subcaption}
\usepackage[font=footnotesize,labelfont=bf]{caption}
\usepackage{lipsum}

\usepackage[symbol]{footmisc}

\newcounter{protocol}
\newenvironment{protocol}[1]
  {\par\addvspace{\topsep}
   \noindent
   \tabularx{\linewidth}{@{} X @{}}
    \hline
    \refstepcounter{protocol}\textbf{Protocol \theprotocol} #1 \\
    \hline}
  { \\
    \hline
   \endtabularx
   \par\addvspace{\topsep}}

\newtheorem{theorem}{Theorem}

\newtheorem{definition}{Definition}

\usepackage{amsmath}

\definecolor{pink}{RGB}{237,16,118}
\definecolor{applegreen}{rgb}{0.55, 0.71, 0.0}
\definecolor{celestialblue}{RGB}{62,146,204}
\definecolor{lilanew}{RGB}{152,41,221}

\newcommand{\COMMENT}[1]{}
\newcommand{\ketbra}[2]{|#1\rangle\langle#2|}

\newcommand{\GHZ}[1]{$\mathrm{GHZ}_{#1}$}

\newcommand{\Fsrhofour}{0.81}

\def\defeq{\mathrel{\mathop:}=} 

\makeatletter
\let\NAT@parse\undefined
\makeatother
\usepackage{hyperref}  

\title{Anonymous Conference Key Agreement \\ in Quantum Networks}

\author[1$ {\ast}$]{Frederik Hahn}
\author[2]{Jarn de Jong}
\author[3,4]{Christopher Thalacker}
\author[3,4]{Bülent Demirel}
\author[3,4]{Stefanie Barz} 
\author[2]{Anna Pappa}
\affil[1]{\emph{Dahlem Center for Complex Quantum Systems, Freie Universit{\"a}t Berlin, 14195 Berlin, Germany}}
\affil[2]{\emph{Electrical Engineering and Computer Science Department, Technische Universit{\"a}t Berlin, 10587 Berlin, Germany}}
\affil[3]{\emph{Institute for Functional Matter and Quantum Technologies, Universit{\"a}t Stuttgart, 70569 Stuttgart, Germany and}}
\affil[4]{\emph{Center for Integrated Quantum Science and Technology (IQST),Universit{\"a}t Stuttgart, 70569 Stuttgart, Germany}}
\affil[$\ast$]{\emph{To whom correspondence should be addressed; E-mail:  frederik.hahn@fulbrightmail.org}}

\begin{document}

\maketitle
\begin{sciabstract}
  Quantum Conference Key Agreement (CKA) is a cryptographic effort of multiple parties to establish a shared secret key. While bipartite quantum key distribution protocols are also useful in the context of CKA, multipartite protocols allow for a more efficient generation of the necessary correlations and are therefore viewed favorably in the context of quantum networks. In future quantum networks, generating secret keys in an anonymous way is of tremendous importance for parties that do not only want to keep their shared key secret but also protect their own identity, e.g. in the context of whistle-blowing.
  
  In this paper we provide the first protocol for Anonymous Quantum Conference Key Agreement and demonstrate it using four-photon Greenberger-Horne-Zeilinger ($\mathbf{GHZ}$) states.
\end{sciabstract}
\clearpage
\section{Introduction}
One of the main applications of quantum information processing is to provide additional security for communication. The most common setting is one of two parties, Alice and Bob, who want to establish a shared secret key in order to encrypt further communication. Since their initial proposal \cite{BB84}, Quantum Key Distribution (QKD) protocols have been proposed and implemented in a standard fashion, even though several practical challenges still remain to be addressed \cite{diamanti_practical_2016}.

In this work, we examine a more generalised scenario, where several parties want to establish a secret key. We introduce a new notion of \emph{anonymity} in this generalised multiparty setting, where we request that the identities of the parties sharing the secret key, are also protected in the best possible way. There are several reasons why such scenarios are highly relevant. One such scenario is the case of whistle-blowing; a person might want to broadcast an encrypted message such that specific parties can decrypt it, while keeping the identities of all involved parties secret. For anonymous whistle-blowing, the underlying protocol needs to involve non-participating parties, such that an authority maintaining the network cannot figure out who takes part in the secret communication. This is, to the best of our knowledge, the first time that anonymity is examined in such a setting, protecting the identity of the sender and of multiple receivers at the same time.

To succeed in attaining our goal, we need to address two different elements, \emph{anonymity} and \emph{multiparty key generation}, often referred to as conference key agreement or CKA (for a concise review, we refer the interested reader to \cite{murta_quantum_2020}). Combining the two, we achieve \emph{anonymous parallel message transmission}, which allows a sender (who we will refer to as Alice) to transmit a private message to specific receivers of her choosing (who we will refer to as Bobs), while keeping their identities secret, both from external parties, and from each other.

Previous work \cite{christandl_quantum_2005} has shown how to achieve anonymous transmission of classical bits using the correlations natural to the \GHZ{} state and also how to anonymously create bipartite entanglement from a larger \GHZ{} state. In \cite{unnikrishnan_anonymity_2019} the latter is developed further, by adding a scheme for anonymous notification of the receiver and a verification scheme \cite{pappa_multipartite_2012, mccutcheon_experimental_2016} to the (anonymous) entanglement generation. Because it is not possible to distill multiple bipartite (e.g. Bell-) states from a single \GHZ{} state, this approach is not sufficient here and we need an alternative approach to be able to perform anonymous CKA over a subset of the entire network. One possibility would be to use other multipartite entangled quantum states  \cite{leung_quantum_2010, hahn_quantum_2019, helwig2013absolutely} to create bipartite entanglement between the sender and all receivers separately. Here, however, we focus on using the \GHZ{n} state shared through the entire network to anonymously establish the necessary entanglement between sender and receivers by using a single quantum state.

In this paper, we introduce a protocol to establish a secret key between Alice as a sender and $m$ receiving parties of her choosing. We use both `Bob' and `receiver' to refer to each of those receiving parties and `participants' to refer to Alice and all the Bobs of her choice. The $m$ Bobs are notified anonymously by Alice through a notification protocol. The $m+1 \leq n$ participants are part of a larger network of $n$ parties. We will start by sharing a large \GHZ{} state between $n$ parties, which can be done either centrally, or using a given network infrastructure via quantum repeaters or quantum network coding \cite{epping_multi-partite_2017}. From the \GHZ{n} state, we subsequently show how to anonymously extract an $\left(m+1\right)$-partite \GHZ{} state shared between Alice and her selection of $m$ Bobs. Repeating the sharing and distillation, the resulting states can either be verified or used to run the CKA protocol. 
\clearpage
\section{Constructing the Anonymous Key Agreement protocol}
We first consider the case where all participants, including the source of entanglement, are honest and trusted. We propose and analyse a protocol for establishing a common secret bit between Alice and $m$ Bobs, while keeping the identities of the $m+1$ participants secret, even from each other. To achieve our goal, we make use of two sub-protocols, which we call \texttt{Notification} and \texttt{Anonymous Multiparty Entanglement}. We now first introduce them separately.

\begin{figure}[h!]
    \centering
    \begin{minipage}[t]{.4\textwidth}
        \centering
        \includegraphics[width=\textwidth]{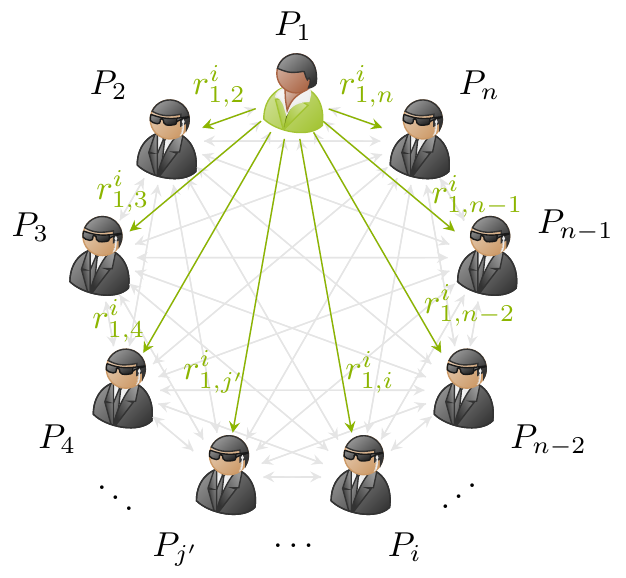}
        \subcaption{Step \ref{Notification_step1a} of \texttt{Notification} with $j_a=1$.}\label{fig:NotificationA}
    \end{minipage}
    \hspace{\baselineskip}
    \begin{minipage}[t]{.4\textwidth}
        \centering
        \includegraphics[width=\textwidth]{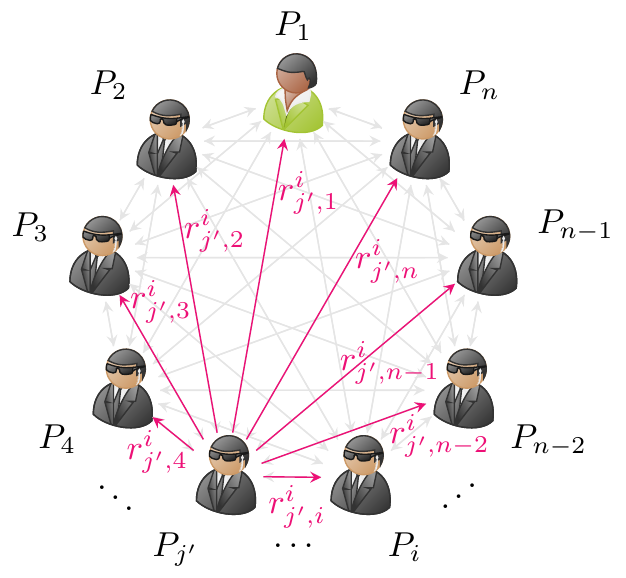}
        \subcaption{Step \ref{Notification_step1b} of \texttt{Notification} with $j=j'$.}\label{fig:NotificationB}
    \end{minipage}
    \\
    \vspace{\baselineskip}
    \begin{minipage}[t]{.4\textwidth}
        \centering
        \includegraphics[width=\textwidth]{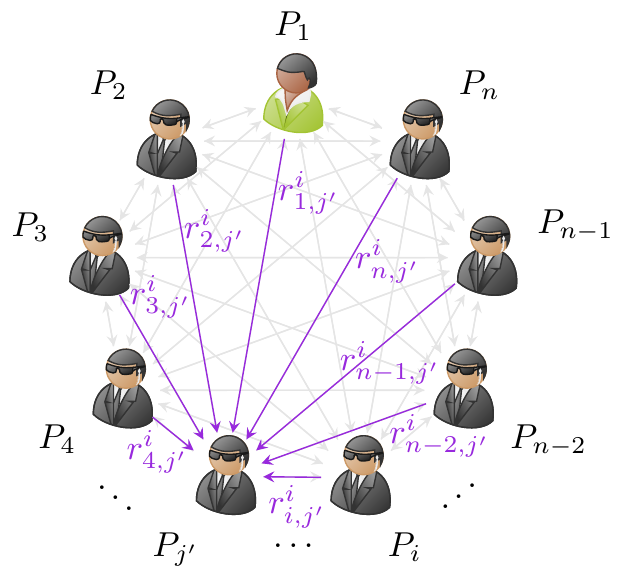}
        \subcaption{Step \ref{Notification_step2} of \texttt{Notification} with $k=j'$.}\label{fig:NotificationC}
    \end{minipage}
    \hspace{\baselineskip}
    \begin{minipage}[t]{.4\textwidth}
        \centering
        \includegraphics[width=\textwidth]{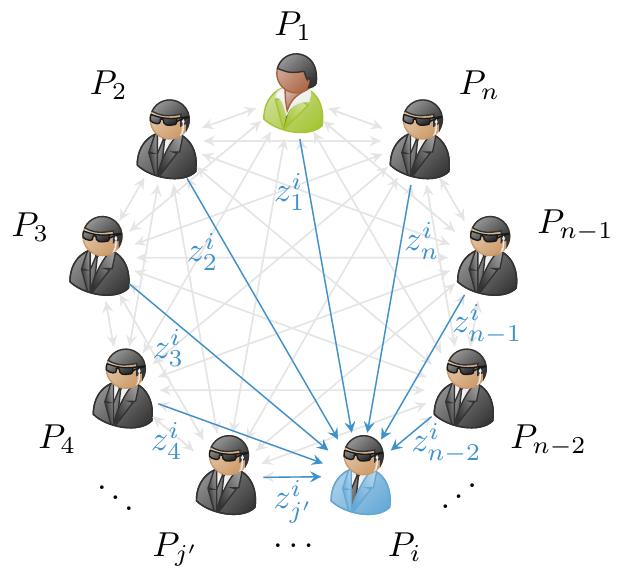}
        \subcaption{Step \ref{Notification_step3} of \texttt{Notification} with $i=i_b$.}\label{fig:NotificationD}
    \end{minipage}
    \\
    \vspace{\baselineskip}
    \includegraphics[width=0.84\textwidth]{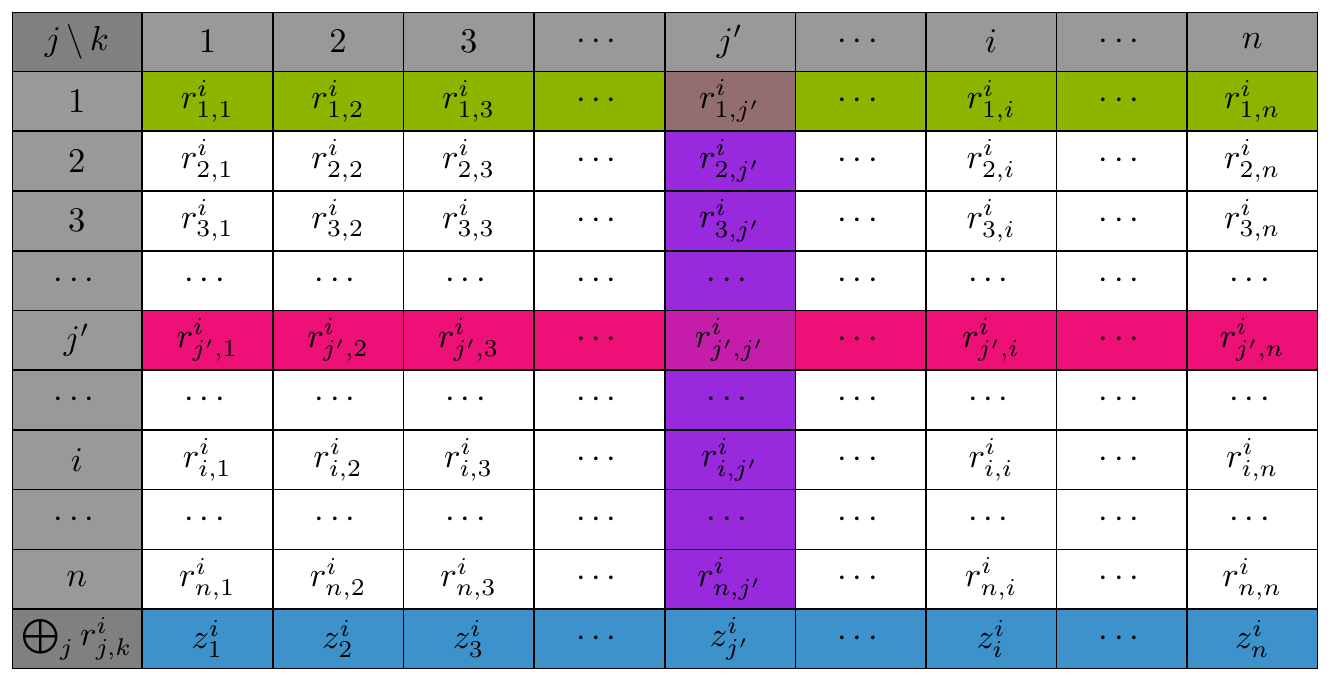}
    \caption{Visualisation of Protocol \ref{Notification}. The table contains all $r^{i}_{j,k}$ for a fixed agent $i\in\{1,\ldots,n\}$ in the \texttt{Notification} protocol. Here, we identify Alice with $P_1$. She chooses $\{r^{i}_{1,k}\}_{k=1}^n$ and sends them to $P_k$ in Step \ref{Notification_step1a} (\textcolor{applegreen}{Fig.\,\ref{fig:NotificationA}}). Note that only if $i$ is a receiver, the \textcolor{applegreen}{green} row adds up to $1 \pmod 2$; otherwise to $0 \pmod 2$. 
	Analogously, the \textcolor{pink}{pink} highlighting shows Step \ref{Notification_step1b} from the perspective of $P_{j'}$ (\textcolor{pink}{Fig.\,\ref{fig:NotificationB}}). This and all other rows add up to $0 \pmod 2$.
	The $\{r^{i}_{j,j'}\}_{j=1}^{n}$ that $P_{j'}$ receives in Step \ref{Notification_step2}  (\textcolor{lilanew}{Fig.\,\ref{fig:NotificationC}}) are highlighted in \textcolor{lilanew}{purple}.   The last row, highlighted in \textcolor{celestialblue}{blue}, shows the $\{z^{i}_{k}\}_{k=1}^{n}$ received by $P_i$ in Step \ref{Notification_step3} (\textcolor{celestialblue}{Fig.\,\ref{fig:NotificationD}}).  By construction, only if $i=i_{b}$  is a receiver, it adds up to $1 \pmod 2$.}
	\label{fig:Notification}
\end{figure}

Our version of \texttt{Notification} is based on \cite{broadbent_information-theoretic_2007} and is a classical protocol used by Alice to notify the $m$ receiving agents, while maintaining anonymity for all parties involved. The protocol requires pairwise private classical communication (which can be established using a key generation protocol based on a Bell pair) and access to  private sources of randomness. An illustration of Protocol \ref{Notification} can be found in Fig.\,\ref{fig:Notification}.
\vspace{0.1in}


\begin{protocol}{\label{Notification} \texttt{Notification}}
\textit{Input.} Alice's choice of $m$ receivers.\\
\textit{Goal.} The $m$ receivers get notified.
\newline\textit{Requirement.} Private pairwise classical communication channels and sources of randomness.

\vspace{\baselineskip}
\noindent For agent $i=1,\dots,n$:
\begin{enumerate}
    \item All agents $j\in\{1,\ldots,n\}$ do the following.\label{Notification_step1}
    \begin{enumerate}
        \item When $j$ corresponds to Alice ($j_a$), and $i$ is not a receiver, she chooses $n$ random bits $\{r_{j,k}^i\}_{k=1}^{n}$ such that $\bigoplus^n_{k=1} r_{j,k}^i = 0$. If $i$ is a receiver, she chooses $n$ random bits such that $\bigoplus^n_{k=1} r_{j,k}^i = 1$. She sends bit $r_{j,k}^i$ to agent $k$ (\textcolor{applegreen}{Fig.\,\ref{fig:NotificationA}}).\label{Notification_step1a}
        \item When $j\neq j_a$, the agent chooses $n$ random bits $\{r_{j,k}^i\}_{k=1}^{n}$ such that $\bigoplus^n_{k=1} r_{j,k}^i = 0$ and sends bit $r_{j,k}^i$ to agent $k$ (\textcolor{pink}{Fig.\,\ref{fig:NotificationB}}).\label{Notification_step1b}
    \end{enumerate}
    \item All agents $k \in \{1,\ldots,n\}$ receive $\{r_{j,k}^{i}\}_{j=1}^{n}$ (\textcolor{lilanew}{Fig.\,\ref{fig:NotificationC}}), compute $z_k^i=\bigoplus_{j=1}^n r_{j,k}^i$ and send it to agent $i$.\label{Notification_step2}
    \item Agent $i$ takes the received $\{z_{k}^{i}\}_{k=1}^{n}$ (\textcolor{celestialblue}{Fig.  \ref{fig:NotificationD}}) to compute $z^i=\bigoplus_{k=1}^n z_k^i$; if $z^i=1$ they are thereby notified to be a designated receiver.\label{Notification_step3}
\end{enumerate}
\end{protocol}\label{Notify}

\paragraph{Analysis:} Anonymity is maintained following the work of \cite{broadbent_information-theoretic_2007}. Remember that by the nature of our goal, the identities of the Bobs are available to Alice since she has chosen them. The \texttt{Notification} protocol requires $\mathcal{O}(n^3)$ communication channel uses between pairs of parties. Note that the \texttt{Notification} protocol is in fact allowing Alice to anonymously communicate a bit to a receiver, and therefore it could in theory be used to share the same bit with all Bobs and thereby establish a common key. Such a process would however be extremely inefficient in the quantum resources, since for each bit of the secret key $\mathcal{O}(n^3)$ Bell pairs would need to be consumed. We could therefore use \texttt{Notification} to expand the preshared randomness that QKD protocols require, but as shown in \cite{epping_multi-partite_2017}, this is less efficient than sharing multipartite entanglement. If instead, we use \texttt{Notification} only once to notify the receivers anonymously, we can exploit the properties of  shared multipartite entangled states to establish a common key more efficiently while maintaining the anonymity that Protocol \ref{Notification} provides.

\clearpage

\begin{figure}[h!]
\begin{centering}
	\includegraphics[width=\textwidth]{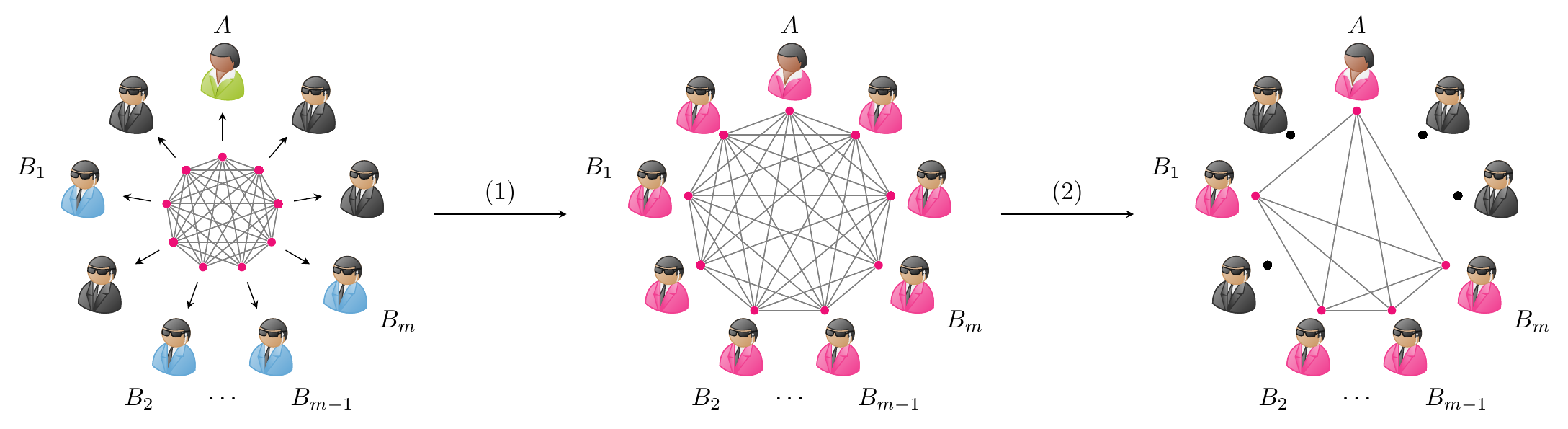}
	\caption{Visualisation of Protocol \ref{AME}. A \GHZ{n} state is shared with all agents left of arrow $(1)$. Here, the participants are highlighted in \textcolor{applegreen}{green} and \textcolor{celestialblue}{blue}. Since the shared \GHZ{n} state is agnostic of the receivers' identities and all agents are entangled right of arrow $(1)$, they are all highlighted in \textcolor{pink}{pink}. Right of arrow $(2)$, all non-participating parties are disentangled and therefore not highlighted anymore. The $m$ Bobs and Alice now share a \GHZ{m+1} state after completing the steps of \texttt{Anonymous Multiparty Entanglement} $(2)$.
	}
	\label{fig:AME}
\end{centering}
\end{figure}

\noindent We now introduce \texttt{Anonymous Multiparty Entanglement}, the second subprotocol. As a generalisation of the protocol first proposed in \cite{christandl_quantum_2005} for anonymously distributing Bell states, it is a protocol for anonymously establishing \GHZ{} states. Here, $n$ parties are sharing a \GHZ{} state, and $m+1$ of them (Alice and $m$ receivers) want to anonymously end up with a smaller, $(m+1)$-partite \GHZ{} state. To achieve this, all parties require access to a broadcast channel -- a necessary requirement to achieve any type of anonymity for the participants in a communication setting \cite{fitzi_unconditional_2002}. Protocol \ref{AME} is visualised in Fig.\,\ref{fig:AME}.

\begin{protocol}{ \texttt{Anonymous Multiparty Entanglement}\label{AME}}
\textit{Input.} A shared \GHZ{} state $\frac{1}{\sqrt{2}}\left(\ket{0}^n+\ket{1}^n\right)$; the $(m+1)$ identities of Alice and the Bobs. \\
\textit{Goal.} An $(m+1)$-partite \GHZ{} state shared between Alice and the $m$ Bobs.
\newline\textit{Requirement.} A broadcast channel; private sources of randomness.
\begin{enumerate}
  \item Alice and the Bobs each draw a random bit. Everyone else 
  measures in the $X$-basis, yielding a measurement outcome bit $x_{i}$.\label{AME_step1}
\item All parties broadcast their bits in a random order or, if possible, simultaneously.\label{AME_step2}
\item Alice applies a $Z$ gate if the parity of the non-participating parties' bits is odd.\label{AME_step3}
  \end{enumerate}
  \end{protocol}
  
\paragraph{Analysis:} The correctness of the protocol follows from the proof in \cite{christandl_quantum_2005}. With the Hadamard matrix $H$ we can rewrite the \GHZ{n} state as

\begin{equation*}
  \frac{1}{\sqrt{2^{n-m}}}\sum_{x\in\{0,1\}^{n-m-1}}\big(\ket{0}_{p}^{m+1}+(-1)^{|x|}\ket{1}_{p}^{m+1}\big)\otimes H^{\otimes(n-m-1)}\ket{x}_{\overline{p}},
\end{equation*}
where $|x|$ is the Hamming weight of $x$ and the subscripts $p$ and $\overline{p}$ indicate the participating and non-participating parties, respectively.
Remember that $H$ interchanges the $X$- and $Z$-bases. After the $X$-measurements of Step \ref{AME_step1}, the state shared between Alice and the Bobs is therefore $\frac{1}{\sqrt{2}}\big(\ket{0}^{m+1}+(-1)^{|x|}\ket{1}^{m+1} \big)$, where $x$ contains all measurement outcomes announced in Step \ref{AME_step2}. Finally, in Step \ref{AME_step3}, Alice can locally correct the state to obtain the desired ($m+1$)-partite \GHZ{} state based on $|x|$.

\noindent With respect to anonymity, the key elements are the intrinsic correlations of the \GHZ{} states. As observed in \cite{christandl_quantum_2005}, any rotation around the $\hat{z}$-axis applied to any of the qubits of a \GHZ{} state has the same effect on the global state independent of the choice of the qubit. To see this, observe that a rotation $R_z(\theta)\defeq (\sigma_z)^{\frac{\theta}{\pi}}$  on any of the $n$ qubits of the \GHZ{} state $\frac{1}{\sqrt{2}}(\ket{0}^n+\ket{1}^n)$ results in the state $\frac{1}{\sqrt{2}}(\ket{0}^n+e^{i\theta}\ket{1}^n)$, therefore not revealing the choice of the qubit. To perform the correction, Alice only needs (the parity of) the measurement outcomes of the non-participating parties, but, to mask their identity, the Bobs announce a (random) bit as well. It is straightforward to see that no one can infer any information about the operations performed by the different parties, since their announced bits are uniformly random, and the application of a $Z$-gate does not reveal the position of the qubit it was applied to to the other parties. Only Alice knows the identities of the Bobs, so only she is able to discern the `true' outcomes from the random bits. Therefore, the protocol does not leak any information about the identity of either Alice or the Bobs.

We can now join the above protocols, in order to achieve  \texttt{Anonymous Key Agreement} between Alice and the $m$ Bobs of her choosing.

\vspace{0.2in}

\begin{protocol}{ \texttt{Anonymous Key Agreement}\label{AKA}}

\textit{Input.} Alice's choice of $m$ Bobs; $L$ \GHZ{n} states. \newline
\textit{Goal.} A common secret key of length $L$ anonymously shared between Alice and the $m$ Bobs.
\newline \textit{Requirement.}  Private pairwise classical communication channels and sources of randomness; a broadcast channel.

\begin{enumerate}
  \item Alice anonymously notifies $m$ chosen Bobs by running the \texttt{Notification} protocol.
  \item All $n$ parties run \texttt{Anonymous Multiparty Entanglement} using the $L$ states.\label{AKA_step2}
  \item Alice and the Bobs $Z$-measure their qubits and obtain a common secret key of length $L$.\label{AKA_step3} \end{enumerate}
  \end{protocol}
 
\vspace{0.1in}
\noindent The above protocol provides two different notions of anonymity, both for the Sender and the Receiver. We define these below.

\begin{definition}[Sender Anonymity]
A protocol allows Alice to remain anonymous sending a message to $m$ Bobs, if an adversary who corrupts $t\leq n - 2$ players, cannot guess the identity of Alice with probability higher than $\frac{1}{n-t}$.
\end{definition}

\begin{definition}[Receiver Anonymity]
A protocol allows Bob to remain anonymous receiving a message from Alice, if an adversary who corrupts $t\leq n - 2$ players, cannot guess the identity of Bob with probability higher than $\frac{1}{n-t}$.
\end{definition}

\noindent It follows directly from the properties of the two subprotocols, that Protocol \ref{AKA} provides both Sender and Receiver Anonymity \cite{broadbent_information-theoretic_2007}, excluding the trivial cases where Alice is corrupted (in which case she already knows the Bobs' identities). Note that this also protects the identities of each Bob and of Alice, even from the other notified participants of the protocol. Note that so far we have considered only honest-but-curious agents; in the next section we will relax this constraint to consider untrusted settings.

\clearpage
\section{Anonymous Key Agreement in untrusted settings}

We will now examine the case were some agents are not honest-but-curious anymore, but are actively trying to extract information from the protocol.

\subsection{Dishonest Source} 
The ostensibly obvious strategy to account for a source that does not share the correct entangled state is to verify the \GHZ{n} state shared by the source directly after sharing \cite{pappa_multipartite_2012, mccutcheon_experimental_2016}.
This approach works for a dishonest source and honest-but-curious agents, since once the state is shared there is no further communication between the parties apart from using the broadcast channel. Therefore, any appropriate verification protocol run after the distillation of the \GHZ{m+1} state (after Step \ref{AKA_step2} of Protocol \ref{AKA}, \texttt{Anonymous Key Agreement}), suffices to detect malicious behavior of the source.
Our \texttt{Verification} protocol is similar to \cite{pappa_multipartite_2012}, and inspired by the pseudotelepathy studies of \cite{brassard_multi-party_2003}, but simplified for the case of honest-but-curious participants and Alice as a fixed verifier.

\vspace{0.2in}

\begin{protocol}{ \texttt{Verification}\label{Ver_GHZ}}
\textit{Input.} A verifier $V$; a shared state between $k$ parties.
\newline
\textit{Goal.} Verification or rejection of the shared state as a \GHZ{k} state by $V$.
\newline
\textit{Requirements.} Private sources of randomness; a classical broadcasting channel.
\begin{enumerate}
  \item Everyone but $V$ draws a random bit $b_i$ and measures in the $X$- or $Y$-basis if their bit equals $0$ or $1$ respectively, obtaining a measurement outcome $m_{i}$.\label{VER_step2}
  \item Everyone broadcasts $(b_i,m_i)$, including $V$, who chooses both at random.
  \item $V$ resets her bit such that $\sum_{i} b_i=0\pmod 2$. She measures in the $X$- or $Y$-basis if her bit equals $0$ or $1$ respectively, thereby also resetting her $m_i=m_v$.\label{VER_step4}
  \item $V$ accepts the state if and only if
  $\sum_{i} m_i=\frac{1}{2}\sum_{i} b_i \pmod{2}.
  $
  \end{enumerate}
  \end{protocol}

\paragraph{Analysis:}
 From \cite{pappa_multipartite_2012}, we know that if the state $\rho$ shared between the parties is far from the \GHZ{} state with respect to the trace distance
\begin{equation*}
d(\rho,\ketbra{\mathrm{GHZ}}{\mathrm{GHZ}})\defeq \frac{1}{2}\left|\mathrm{tr}(\rho-\ketbra{\mathrm{GHZ}}{\mathrm{GHZ}})\right|,
\end{equation*}
 then $V$ will reject the state with high probability. Denoting by $\text{T}(\rho)=1$ the event that the \texttt{Verification} protocol accepts a state with density matrix $\rho$, the following theorem holds.
\begin{theorem}[cf. \cite{pappa_multipartite_2012}]\label{theorem_1}
If d$(\rho,\ketbra{\mathrm{GHZ}}{\mathrm{GHZ}})=\epsilon$, then $\Pr[\text{T}(\rho)=1]\leq 1-\frac{\epsilon^2}{2}$.
\end{theorem}

\subsection{Dishonest agents}
In the more general case where the agents can also be malicious however, any verification technique used on the large \GHZ{} state (like protocol \ref{Ver_GHZ}, also used in \cite{unnikrishnan_anonymity_2019}) creates a critical security problem. An agent not belonging to the set of designated receivers, might not measure in Step \ref{AME_step1} of Protocol \ref{AME}, and thereby, when it is used during Protocol \ref{AKA}, end up sharing a \GHZ{} state with the legitimate participants. This security risk was independently noticed in \cite{yang_examining_2020} for the case of two-party communication.

It is now evident that in the setting of dishonest agents the \texttt{Verification} protocol \textit{has} to be performed after distilling the \GHZ{m+1} state, whereas for the setting of (only) a dishonest source this was not necessary but only preferred.
We are thus required to `postpone' the verification step till after Step \ref{AKA_step2} of Protocol \ref{AKA}, to make sure that only Alice and the chosen Bobs share the final \GHZ{} state. Our proposed process keeps their identities secret, while they apply a verification protocol similar to the one presented above for verifying a \GHZ{} state. We are now ready to define Protocol \ref{AKA_ver1} for anonymously sharing a key between Alice and $m$ Bobs, where $L$ is the number of shared GHZ-states and $D$ is a parameter both determining the level of security and the length of the generated shared key. The main difference between the proposed protocol and the one in \cite{unnikrishnan_anonymity_2019}, is that the non-participating parties are asked to announce random values, and that the protocol aborts if the values are not announced in time.

\begin{protocol}{ \texttt{Anonymous Verifiable Key Agreement}\label{AKA_ver1}}
\textit{Input.} Alice as the verifier; parameters $L$ and $D$. 
\newline
\textit{Goal.} Anonymous generation of secret key between Alice and $m$ Bobs.
\newline
\textit{Requirements.} A source of \GHZ{n} states; private sources of randomness; a random source that is not associated with any party; a classical broadcasting channel; pairwise private classical communication channels.

\begin{enumerate}
  \item Alice notifies the $m$ Bobs by running the \texttt{Notification} protocol.\label{AVKA_Notification}
  \item The source generates and shares $L$ \GHZ{} states.\label{AVKA_L_GHZ_states}
  \item The parties run the \texttt{Anonymous Multiparty Entanglement} protocol on them.
  \item The parties ask a source of randomness to broadcast a bit $b$ such that $\Pr[b=1]=\frac{1}{D}$.
  \begin{description}
   \item[\texttt{Verification} round:] If $b=0$, Alice runs the \texttt{Verification} protocol on the $(m+1)$-partite  state, therefore only considering the announcements of the $m$ Bobs. The remaining parties announce random values. 
  \item[\texttt{KeyGen} round:] If $b=1$, Alice and the Bobs measure in the $Z$-basis .
\end{description}
\item If Alice is content with the checks of the \texttt{Verification} protocol, she can anonymously validate the protocol.
\end{enumerate}
  \end{protocol}

\paragraph{Analysis:} The above protocol aims to establish a secret key between Alice and $m$ Bobs, while keeping their identities secret, both from outsiders and each other. The correlations between the inputs and the outputs of Alice and the Bobs, can only be observed by Alice, since they look random to anyone but Alice. In addition, when Alice is verifying the shared \GHZ{} state between her and the $m$ Bobs of her choice, she can indirectly verify that the \texttt{Notification} protocol has run correctly, since otherwise some of them would not be able to provide the appropriate input/output correlations when running the \texttt{Verification} protocol. As the latter protocol verifies that the state shared between Alice and the $m$ Bobs is close to the \GHZ{} state, and by default they are all honest, Theorem \ref{theorem_1} holds. If Alice accepts the checks of the \texttt{Verification} protocol, anonymity is also maintained since the state is close to the \GHZ{} state (except with some small probability). 

On average $D-1$ states will be used to verify the state and only one to provide a secret key. Therefore the key rate of Protocol \ref{AKA_ver1} approaches  $\frac{L}{D}$ in the asymptotic regime.

\section{Experimental Implementation}
\begin{figure}[h!]
    \centering
    \includegraphics[width=0.4\textwidth]{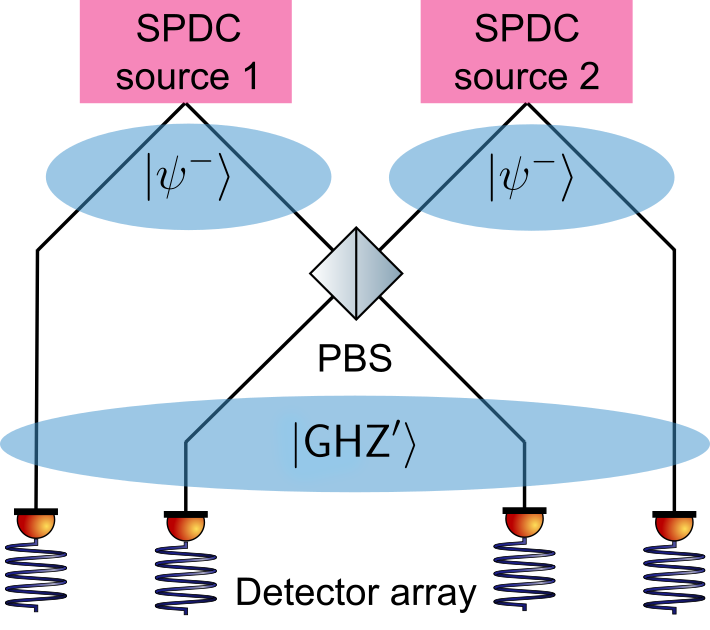}
    \caption{A laser pumps two SPDCs after which both emit a photon pair. The state of each photon pair can be described as $\ket{\psi^-}=\frac{1}{\sqrt{2}}(\ket{H,V}-\ket{V,H})$ ($H=0,  V=1$). Subsequently, two photons, one photon from each pair, interfere in a polarizing beam splitter (PBS), after which post-selecting only four-photon events the state $\ket{\mathrm{GHZ}'}=\frac{1}{\sqrt{2}}(\ket{H,V,V,H}-\ket{V,H,H,V})$ is obtained; this state is locally equivalent to the \GHZ{4} state. The state fidelity is calculated to be $F = \Fsrhofour{}$.}
    \label{fig:setup}
\end{figure}

To test the \texttt{Anonymous Verifiable Key Agreement} protocol experimentally we
de-monstrate the \textbf{\texttt{Verification}} and \textbf{\texttt{KeyGen} rounds} experimentally.
For simplicity we omit running Step \ref{AVKA_Notification} of the protocol, since this can be done by standard implementations of BB$84$ links. We request our source to prepare $L$ \GHZ{4} states on which we run the \texttt{Verification} protocol using a random seed. 
Our demonstration uses a four-photon \GHZ{} state in polarisation encoding ($H=0,  V=1$) generated using two parametric down-conversion sources \cite{demirel2020correlations}. The setup is displayed in Fig.\,\ref{fig:setup}.

\noindent Without loss of generality we can assume that a fixed party always plays the role of Alice who wants to obtain a common key with her choice of Bobs. We will consider all three configurations of her choosing two Bobs to establish this common key while preserving the anonymity of all participants. 
Tab.\,\ref{tab:Configurations} shows the different measurement operators used in the different configurations. 
\begin{table}[h]
\centering
\begin{tikzpicture}
\matrix (M) [
    matrix of nodes,
    minimum width=1cm,
    minimum height=0.7cm,
    column sep=0mm,
    row sep=0mm,
    nodes={
        draw,
        line width=0.07mm,
        anchor=south,
        align=center, 
    },
    row 1/.style={
        nodes={
           minimum height=0.7cm,
        }
    },
    row 7/.style={
        nodes={
           minimum height=0.7cm
        }
    },
    column 1/.style={
        nodes={
            text width=1.618*3cm,
            minimum width=1cm,
            fill=gray
        }
    },
    column 2/.style={
        nodes={
            text width=1.618cm,
            minimum width=1cm,
        }
    },
    column 3/.style={
        nodes={
            text width=1.618cm,
            minimum width=1cm,
            fill=gray
        }
    },
    column 4/.style={
        nodes={
            text width=1.618cm,
            minimum width=1cm,
            fill=gray
        }
    }        
]{  
|[fill=gray]| Configuration
&|[fill=pink]|$AB_1B_2P_4$ 
&|[fill=applegreen]|$AP_2B_1B_2$ 
&|[fill=celestialblue]|$AB_1P_3B_2$\\\\
|[fill=gray!80]|
\textbf{\texttt{Verification}} $(0,0,0)$
    &|[fill=pink!80]| 
    $XXXX$ 
    &|[fill=applegreen!80]| 
    $XXXX$ 
    &|[fill=celestialblue!80]| 
    $XXXX$\\
|[fill=gray!60]|
\textbf{\texttt{Verification}}
$(0,1,1)$
    &|[fill=pink!60]| 
    $XYYX$ 
    &|[fill=applegreen!60]| 
    $XXYY$ 
    &|[fill=celestialblue!60]| 
    $XYXY$\\
|[fill=gray!40]|
\textbf{\texttt{Verification}}
$(1,0,1)$
    &|[fill=pink!40]| 
    $YXYX$ 
    &|[fill=applegreen!40]| 
    $YXXY$ 
    &|[fill=celestialblue!40]| 
    $YXXY$ \\
|[fill=gray!20]|
\textbf{\texttt{Verification}} 
$(1,1,0)$ 
    &|[fill=pink!20]| 
    $YYXX$
    &|[fill=applegreen!20]| 
    $YXYX$ 
    &|[fill=celestialblue!20]| 
    $YYXX$\\
|[fill=gray]|\textbf{\texttt{KeyGen}} 
    &|[fill=pink]| 
    $ZZZX$
    &|[fill=applegreen]| 
    $ZXZZ$ 
    &|[fill=celestialblue]| 
    $ZZXZ$ \\
};
\end{tikzpicture}
\caption{The last row describes the \textbf{\texttt{KeyGen} rounds}; rows two to five describe all possible \textbf{\texttt{Verification} rounds}  $(b_1,b_2,b_3)$ depending on the randomly drawn bits $b_i$. Alice resetting her bit in Step \ref{VER_step4} of Protocol \ref{Ver_GHZ} corresponds to having an even number of $Y$-measurements in every \textbf{\texttt{Verification} round}.}
\label{tab:Configurations}
\end{table}

\begin{figure}
\begin{centering}
	\includegraphics[width=0.9\textwidth]{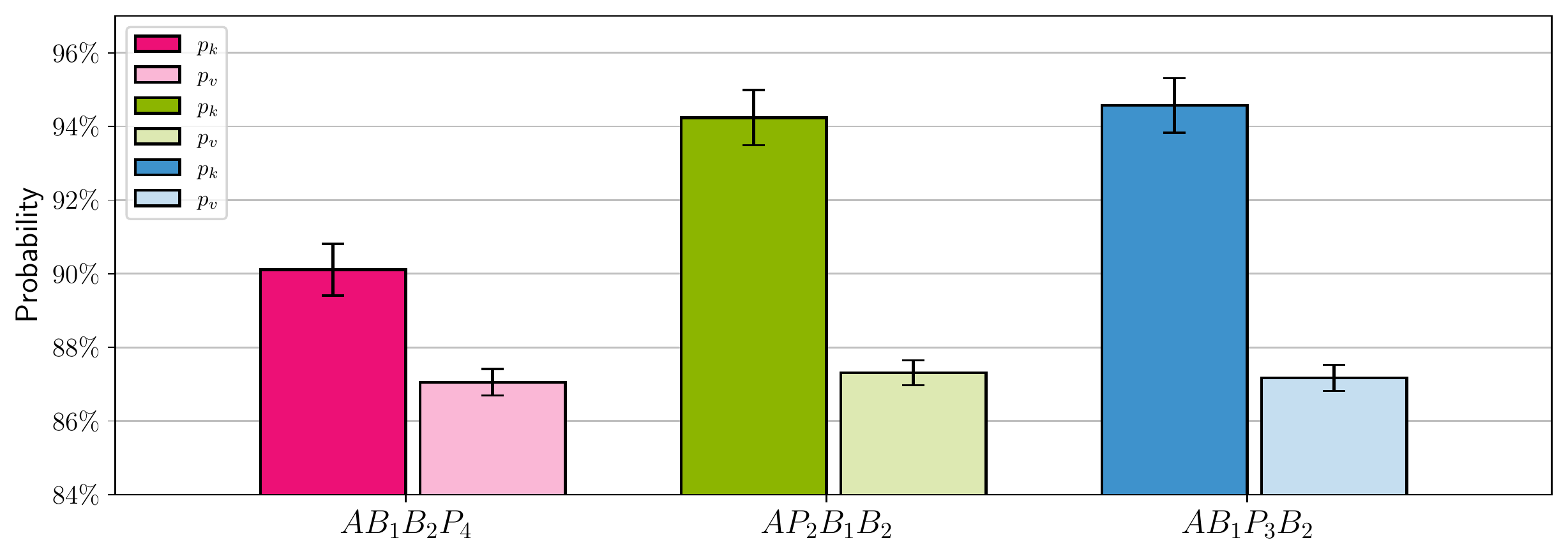}
	\caption{Probabilities of a successful \textbf{\texttt{KeyGen} round} $p_{k}$ (left) and \textbf{\texttt{Verification} round} $p_{v}$ (right), for all three network configurations listed in Tab.\,\ref{tab:Configurations}. The probabilities are calculated as the number of correct measurements divided by the total number of measurements; each $p_{v}$ is the average of the four different measurement operators for that specific configuration.}
	\label{fig:keyverprobabilities}
\end{centering}
\end{figure}

Fig.\,\ref{fig:keyverprobabilities} shows the probabilities of a successful \textbf{\texttt{Keygen} round} $p_{k}$ and \textbf{\texttt{Verification} round} $ p_{v}$ for all three different network configurations. From this, we can calculate the overall probabilities averaged over all network configurations to be $\hat{p}_{k} = 92.974 \pm 0.4230$ and $\hat{p}_{v} = 87.178 \pm 0.2028$. 



\newpage
\section{Discussion and conclusion}
In this work, we demonstrated how to establish a common key between several parties, while keeping their identities secret. Studies of anonymous entanglement have been conducted based on various quantum states, that share a different type of entanglement than the \GHZ{} state; for instance in \cite{lipinska_anonymous_2018} anonymous transmissions using $\textrm{W}$ states is studied. However, we specifically focus on \GHZ{} states since they
show straightforward correlations that can be used to achieve key agreement. If we want to correct for errors and maximise the secret key rate, we need to perform Error Correction and Privacy Amplification schemes \cite{epping_multi-partite_2017, proietti_experimental_2020} while preserving anonymity. This can be quite intricate and should therefore be carried out carefully. We leave this as an open question to be addressed in subsequent work.

\end{document}